\pdfoutput=1
\documentclass{article}

     \usepackage[preprint,nonatbib]{neurips_2021}

\usepackage[utf8]{inputenc} 
\usepackage[T1]{fontenc}    
\usepackage{hyperref}       
\usepackage{url}            
\usepackage{booktabs}       
\usepackage{amsfonts}       
\usepackage{nicefrac}       
\usepackage{microtype}      
\usepackage{bm}
\usepackage{physics}
\usepackage{mathtools}
\mathtoolsset{showonlyrefs}
\usepackage{xcolor}
\usepackage{subfigure}
\usepackage{comment}

\usepackage{algorithmic}
\usepackage[linesnumbered,ruled,noend]{algorithm2e} 

\usepackage{multirow}

\newtheorem{remark}{Remark}[section]

\newcommand{\Transpose}{^{\mathsf{T}}}

\newcommand{\PoissonBracket}[2]{\{#1,#2\}}
\newcommand{\PoissonBracketBi}[2]{\pdv{#1}{\State}\PoissonMatrix\pdv{#2}{\State}}

\newcommand{\PoissonMatrix}{L}

\newcommand{\PoissonMatrixNN}{\PoissonMatrix_{\PoissonNNParam}}

\newcommand{\IrrBracket}[2]{[#1,#2]}
\newcommand{\IrrBracketBi}[2]{\pdv{#1}{\State}\FrictionMatrix\pdv{#2}{\State}}
\newcommand{\FrictionMatrix}{M}
\newcommand{\FrictionMatrixNN}{\FrictionMatrix_{\FrictionNNParam}}

\newcommand{\TotalEnergy}{E}
\newcommand{\TotalEnergyArg}[1]{\TotalEnergy(#1)}
\newcommand{\TotalEnergyNN}{\TotalEnergy_{\EnergyNNParam}}
\newcommand{\TotalEnergyNNArg}[1]{\TotalEnergyNN(#1)}

\newcommand{\Entropy}{S}
\newcommand{\EntropyArg}[1]{\Entropy(#1)}
\newcommand{\EntropyNN}{\Entropy_{\EntropyNNParam}}
\newcommand{\EntropyNNArg}[1]{\EntropyNN(#1)}

\newcommand{\Hamiltonian}{H}

\newcommand{\EnergyNNParam}{\bm{\phi}}
\newcommand{\EntropyNNParam}{\bm{\varphi}}
\newcommand{\FrictionNNParam}{\bm{\rho}}
\newcommand{\PoissonNNParam}{\bm{\pi}}

\newcommand{\State}{\bm{x}}

\newcommand{\ApproxState}{\tilde{\State}}
\newcommand{\TimeSymb}{t}
\newcommand{\FinalTime}{\TimeSymb_{\mathrm{final}}}
\newcommand{\TrainTime}{\TimeSymb_{\mathrm{train}}}
\newcommand{\ValTime}{\TimeSymb_{\mathrm{val}}}
\newcommand{\TestTime}{\TimeSymb_{\mathrm{test}}}
\newcommand{\TimeStep}{\Delta \TimeSymb}

\newcommand{\PositionSymb}{q}
\newcommand{\MomentumSymb}{p}

\newcommand{\InitState}{\State^{0}}

\newcommand{\LossSymb}{\mathcal L}

\newcommand{\DynamicsFunction}{\bm{f}}
\newcommand{\NNParams}{\Theta}
\newcommand{\DynamicsNN}{\DynamicsFunction_{\NNParams}}

\newcommand{\NNParamsNODE}{\theta}
\newcommand{\DynamicsNODE}{\DynamicsFunction_{\NNParamsNODE}}

\newcommand{\Observable}{\State^{\mathrm{o}}}
\newcommand{\ApproxObservable}{\tilde{\State}^{\mathrm{o}}}

\newcommand{\Unobservable}{\State^{\mathrm{u}}}
\newcommand{\ApproxUnobservable}{\tilde{\State}^{\mathrm{u}}}

\newcommand{\batchlength}{L}
\newcommand{\batchsize}{N_{\mathrm{b}}}

\newcommand{\maxepoch}{n_{\max}}
\newcommand{\nupdate}{n_{\mathrm{update}}}

\title{Machine learning structure preserving brackets for forecasting irreversible processes}

\author{%
   Kookjin Lee\\
   School of Computing, Informatics,\\ and Decision Systems Engineering \\
   Arizona State University\\
   Tempe, AZ 85281 \\
   \And  
   Nathaniel A. Trask \\
  Center for Computing Research \\
  Sandia National Laboratories \\
  Albuquerque, NM 87123 \\
  \texttt{natrask@sandia.gov} \\
  \And
  Panos Stinis\\
  Pacific Northwest National Laboratory\\
  Richland, WA 99354
}

\begin{document}

\maketitle

\begin{abstract}
  Forecasting of time-series data requires imposition of inductive biases to obtain predictive extrapolation, and recent works have imposed Hamiltonian/Lagrangian form to preserve structure for systems with \emph{reversible} dynamics. In this work we present a novel parameterization of dissipative brackets from metriplectic dynamical systems appropriate for learning \emph{irreversible} dynamics with unknown a priori model form. The process learns generalized Casimirs for energy and entropy guaranteed to be conserved and nondecreasing, respectively. Furthermore, for the case of added thermal noise, we guarantee exact preservation of a fluctuation-dissipation theorem, ensuring thermodynamic consistency. We provide benchmarks for dissipative systems demonstrating learned dynamics are more robust and generalize better than either "black-box" or penalty-based approaches.
\end{abstract}

\section{Background and previous work}

Modeling time-series data as a solution to a dynamical system with learnable dynamics has been shown to be effective in both data-driven modeling for physical systems and traditional machine learning (ML) tasks. Broadly, it has been observed that imposition of physics-based structure leads to more robust architectures which generalize well \cite{baker2019workshop}. On one end of the spectrum of inductive biases, universal differential equations (UDE) \cite{rackauckas2020universal} assume an a priori known model form, thus imposing the strongest bias. On the other, neural ordinary differential equations (NODEs) \cite{chen2018neural} assume a completely black-box model form with minimal bias. 

Many recent approaches have turned to structure preserving models of reversible dynamics to obtain an inductive bias that lies in between \cite{Greydanus2019hnn,cranmer2020lagrangian,chen2019symplectic,jin2020sympnets,tong2021symplectic}. One may use black-box deep neural networks (DNNs) to learn an energy of a system with unknown model form, while the algebraic structure of Hamiltonian/Lagrangian dynamics provides a flow map which conserves energy. Typically, the learned flow map has symplectic structure so that phase space trajectories are conserved. In classification problems, this mitigates the vanishing/exploding gradient problem and improves accuracy \cite{haber2017stable}; in physics, this guarantees that extrapolated states are physically realizable \cite{sanchez2019hamiltonian}.

Such approaches are only appropriate for reversible systems lacking friction or dissipation. In the physics literature, the theory of metriplectic dynamical systems provides a generalization of the Poisson brackets of Hamiltonian/Lagrangian mechanics which model not just a conserved energy, but generalized Casimirs such as entropy \cite{guha2007metriplectic,grmela1997dynamics}. Physical systems which can be cast in this framework obtain a number of mimetic properties related to thermodynamic consistency: satisfaction of the first and second laws of thermodynamics and, for closed stochastic systems, a fluctuation dissipation theorem (FDT) that guarantees thermal forcing is balanced exactly by dissipative forces in equilibrium \cite{ottinger2020framework}. This FDT property is particularly critical to analyzing rare events in molecular phenomena driven by thermal noise \cite{montefusco2018coarse}. 

Classically, metriplectic systems are obtained by deriving a model from first principles and then observing that the system admits a requisite algebraic structure. While effective for a wide range of physical systems \cite{ottinger2018generic}, the requisite first principles modeling may be restrictively complex for a general system, particularly for multiscale problems involving time history. \emph{In this work we reverse the process by assuming our time-series data has been generated by a metriplectic system and then inferring the requisite algebraic objects using a training strategy similar to that used in NODEs}. This presents several technical challenges. First, dissipative systems typically have non-observable states (i.e. internal entropy or temperature) which may not be measured. Second, the metriplectic algebraic structure is particularly restrictive, requiring discovery of matrices with carefully designed null-spaces to separate reversible and irreversible components of the dynamics.


\textbf{Anticipated impact:} Finally, we note that this work is an important first step toward handling more complicated dissipative chaotic systems ubiquitous to science and engineering problems. For example, for chaotic systems the ``butterfly effect'' causes arbitrarily small perturbations in initial data to exponentially diverge, and it is only possible to provide long-term forecasting by learning a corresponding ``strange attractor'' whose latent dimension is governed by the dissipative structure \cite{lorenz1963deterministic,grassberger1983characterization}. In reduced order-modeling, many have looked toward data-driven means of fitting dynamics to latent representations of solution space, with Hamiltonian structure particularly useful for finding long-time accurate surrogates\cite{gao2021non,hesthaven2018structure}. In this situation as well, structure-preserving treatment of dissipation is critical to account for entropic/memory effects which emerge from coarse-graining \cite{chorin2007problem}. Another success of reversible structure-preserving ML is in robotic control \cite{lutter2019deep}. Again these models fail to account for friction due to wear, which is inevitable in realistic applications.

\section{Related work}
\paragraph{Neural ordinary differential equations} 
As noted, learning time-continuous dynamics in the form of a system of ODEs is an active topic with seminal works including \cite{weinan2017proposal,haber2017stable,lu2018beyond,chen2018neural, NEURIPS2019_42a6845a}. There have been many follow-up studies to enhance neural ODEs in different aspects, e.g., enhancing the expressivity of neural ODEs by augmenting extra dimensions in state variables \cite{dupont2019anode}, checkpoint methods to mitigate numerical instability and to enhance memory efficiency \cite{gholami2019anode,zhuang2020adaptive,zhuang2021mali}, allowing network parameters to evolve over time together with hidden states \cite{zhang2019anodev2,massaroli2020dissecting}, and spectrally approximating dynamics by using a set of orthogonal polynomials \cite{quaglino2019snode}. Applications of NODEs for learning complex physical processes (e.g.,  turbulent flow) can be found in \cite{maulik2020time,portwood2019turbulence,lee2020parameterized}.

\paragraph{Structure preserving neural networks} A thorough accounting of works embedding structure-preservation into neural networks include pioneering works for Hamiltonian neural networks \cite{Greydanus2019hnn,toth2019hamiltonian}, followed by development of Lagragian neural networks \cite{lutter2018deep,cranmer2020lagrangian} and neural networks that mimic the action of symplectic integrators \cite{chen2019symplectic,jin2020sympnets,tong2021symplectic}. More recently, there has been efforts to add physical invariance to learned dynamics models, e.g., time-reversal symmetry \cite{huh2020time}. Works pursuing related but distinct spatial-compatibility related to conservation structure other than geometric integration include: graph architectures with associated a data-driven graph exterior calculus \cite{trask2020enforcing}, solving optimization problems with conservation constraint in latent space \cite{lee2019deep}, and adding conservation constraints as a penalty in training loss \cite{beucler2019achieving}. The closest work to our approach is in \cite{hernandez2021structure}, which proposed a time integrator that leverages the GENERIC (general equation for the nonequilibrium reversible–irreversible coupling) formalism to impose the structure, but enforces the degeneracy condition as penalty terms in the training loss objective. We will provide results demonstrating that a penalty approach is insufficient to guarantee preservation of metriplectic structure.

\section{Theory and fundamentals}\label{sec:theory}

We consider the GENERIC formalism as a particular metriplectic framework amenable to parameterization. Consider time series data $\mathcal{D} = \left\{\left(t_i,\State(t_i)\right)\right\}_{i=1}^N$, where the state $\State_i = \State(t_i)\subset \mathbb{R}^d$ has a known initial condition $\State_0$. In GENERIC, it is assumed that an observable $A(\State)$ evolves under the gradient flow
\begin{equation}\label{eq:generic_bracket}
    \dv{A}{\TimeSymb} = \PoissonBracket{A}{\TotalEnergy} + \IrrBracket{A}{\Entropy}
\end{equation}
where $\TotalEnergy$ and $\Entropy$ denote generalized energy and entropy, $\PoissonBracket{\cdot}{\cdot}$ denotes a Poisson bracket, and $\IrrBracket{\cdot}{\cdot}$ denotes an irreversible bracket. The Poisson bracket is given in terms of a skew-symmetric Poisson matrix $\PoissonMatrix$ and the irreversible bracket is given in terms of a symmetric positive semi-definite friction matrix $\FrictionMatrix$,
\begin{equation}\label{eq:brackets}
    \PoissonBracket{A}{B} = \PoissonBracketBi{A}{B},\quad \mathrm{and} \quad \IrrBracket{A}{B} = \IrrBracketBi{A}{B}.
\end{equation}
A system governed by Eq.~\ref{eq:generic_bracket} is a GENERIC system if the following degeneracy conditions hold
\begin{equation}\label{eq:degeneracy_cond}
    \PoissonMatrix \pdv{\Entropy}{\State} = 0, \quad \mathrm{and} \quad \FrictionMatrix \pdv{\TotalEnergy}{\State}=0.
\end{equation}

Taking $A=\State$ in Eq.~\eqref{eq:generic_bracket} provides the evolution of $\State$
\begin{equation}\label{eq:generic}
    \dv{\State}{\TimeSymb} = \PoissonMatrix \pdv{\TotalEnergy}{\State} + \FrictionMatrix\pdv{\Entropy}{\State}.
\end{equation}

\begin{remark}[Hamiltonian dynamics]
For canonical coordinates $\State = [\PositionSymb,\MomentumSymb]\Transpose$, and canonical Poisson matrix $\PoissonMatrix= \begin{bmatrix}0 & 1 \\ -1 & 0\end{bmatrix}$, and $\FrictionMatrix = \bm{0}$, Eq.~\eqref{eq:generic} recovers Hamiltonian dynamics.
\end{remark}

\begin{remark}[First and second laws of thermodynamics]
Taking $A=E$ and $A=S$, we obtain $\dv{E}{\TimeSymb}=0$ and $\dv{S}{\TimeSymb}\geq 0$, respectively. This follows easily by application of the degeneracy conditions and noticing $\PoissonBracket{A}{A}=0$, $\IrrBracket{A}{A}\geq 0$. 
\end{remark}
\begin{remark}[Fluctuation dissipation theorem]
Introducing thermal noise to Eq.~\eqref{eq:generic} provides the stochastic differential equation (SDE)
\begin{equation}\label{eq:genericSDE}
    \mathrm{d} \State_t = \left(\PoissonMatrix \pdv{\TotalEnergy}{\State} + \FrictionMatrix\pdv{\Entropy}{\State} + k_B \pdv{}{\State} \cdot \FrictionMatrix\right) \mathrm{d}t + \sqrt{ 2 k_B\FrictionMatrix}\mathrm{d}W_t,
\end{equation}
where $\sqrt{\FrictionMatrix}$ denotes the Cholesky factor of $\FrictionMatrix$, $k_B$ is a Boltzmann constant, and $\mathrm{d}W_t$ is a Wiener process. The equilibrium statistics of this SDE reach a stationary distribution under appropriate conditions \cite{ottinger2020framework}. 
\end{remark}

\section{Parameterization of bracket structure}\label{sec:param_generic}
We now introduce a parameterization of the dissipative and reversible brackets that exactly satisfies the degeneracy conditions described in Section \ref{sec:theory}, and review the penalty approach from \cite{hernandez2021structure} which imposes degeneracy conditions via soft constraints. Our approach is motivated by the work in \cite{oettinger2014irreversible} which we summarize in Sections \ref{sec:rev}--\ref{sec:degen}. For the remainder, we adopt the Einstein summation convention.  

First, we parameterize the energy and the entropy as neural networks, i.e., $\TotalEnergyArg{\State} \approx \TotalEnergyNNArg{\State}$ and $\EntropyArg{\State} \approx \EntropyNNArg{\State}$, where $\EnergyNNParam$ and $\EntropyNNParam$ are weights and biases for the neural networks $E$ and $S$ respectively.

\subsection{Parameterizing skew-symmetric reversible dynamics}\label{sec:rev}
The reversible dynamics are characterized by a skew-symmetric Poisson bracket, 
\begin{equation}
    \PoissonBracket{A}{B} = \xi_{\alpha\beta\gamma} \pdv{A}{x_\alpha} \pdv{B}{x_\beta} \pdv{\Entropy}{x_\gamma},
\end{equation}
where $\xi_{\alpha\beta\gamma}$ is an skew-symmetric 3d tensor. To enforce the anti-symmetry exactly, we consider a generic 3 tensor $\tilde{\xi}_{\alpha \beta \gamma}$ with learnable entries and apply the following skew-symmetrization trick.
\begin{equation}
    \xi_{\alpha\beta\gamma} = \frac{1}{3!}\left( \tilde{\xi}_{\alpha\beta\gamma} - \tilde{\xi}_{\alpha\gamma\beta} + \tilde{\xi}_{\beta\gamma\alpha} - \tilde{\xi}_{\beta\alpha\gamma} + \tilde{\xi}_{\gamma\alpha\beta} - \tilde{\xi}_{\gamma\beta\alpha} \right).
\end{equation}

The reversible part may then be written as $ \{\State,\TotalEnergy\} = \xi_{\alpha\beta\gamma} \pdv{\State}{x_\alpha} \pdv{\TotalEnergy}{x_\beta} \pdv{\Entropy}{x_\gamma}$ and the reversible dynamics are given by 
\begin{equation}
    \left(\dv{x_\alpha}{t}\right)_{\mathrm{r}} = \xi_{\alpha\beta\gamma}  \pdv{\TotalEnergy}{x_\beta} \pdv{\Entropy}{x_\gamma}. 
\end{equation}

\subsection{Parameterizing symmetric irreversible dynamics}
Next, we parameterize the irreversible dynamics via the bracket, 
\begin{equation}
    \IrrBracket{A}{B} = \zeta_{\alpha\beta,\mu\nu} \pdv{A}{x_\alpha} \pdv{\TotalEnergy}{x_\beta} \pdv{B}{x_\mu} \pdv{\TotalEnergy}{x_\nu},
\end{equation}
where
\begin{equation}
    \zeta_{\alpha\beta,\mu\nu} = \Lambda_{\alpha\beta}^m D_{mn} \Lambda_{\mu\nu}^n.
\end{equation}
Here, $\Lambda$ and $D$ are skew-symmetric and symmetric positive semi-definite matrices, respectively, such that
\begin{equation}
    \Lambda_{\alpha\beta}^m = -\Lambda_{\beta\alpha}^m, 
    \quad \mathrm{and} \quad
    D_{mn} = D_{nm}.
\end{equation}
Again, the skew-symmetry and the symmetric positive semi-definiteness can be achieved by the parameterization tricks
\begin{equation}
    \Lambda = \frac{1}{2}(\tilde \Lambda - \tilde \Lambda\Transpose), \quad \mathrm{and} \quad \quad D = \tilde D \tilde D\Transpose,
\end{equation}
where $\tilde{\Lambda}$ and $\tilde{D}$ are matrices with learnable entries.
Finally, the irreversible part may be written as $\IrrBracket{\State}{\Entropy} = \zeta_{\alpha\beta,\mu\nu} \pdv{\State}{x_\alpha} \pdv{\TotalEnergy}{x_\beta} \pdv{\Entropy}{x_\mu} \pdv{\TotalEnergy}{x_\nu}$ and the irreversible part of the dynamics is given by
\begin{equation}
    \left(\dv{x_\alpha}{t}\right)_{\mathrm{irr}} = \zeta_{\alpha\beta,\mu\nu} \pdv{\TotalEnergy}{x_\beta} \pdv{\Entropy}{x_\mu} \pdv{\TotalEnergy}{x_\nu}.
\end{equation}

\subsection{Degeneracy conditions}\label{sec:degen}
With the above parameterizations the degeneracy conditions described in Eq.~\eqref{eq:degeneracy_cond} may be easily verified by direct calculation following the definition of the brackets and the symmetry/skew-symmetry conditions. 
\begin{equation}
    \PoissonBracket{\State}{\Entropy} = \pdv{\State}{\State} \PoissonMatrix \pdv{\Entropy}{\State} = \xi_{\alpha\beta\gamma} \pdv{\State}{x_\alpha} \pdv{\Entropy}{x_\beta} \pdv{\Entropy}{x_\gamma} = \xi_{\alpha\beta\gamma} \pdv{\Entropy}{x_\beta} \pdv{\Entropy}{x_\gamma} =0,
\end{equation}
and 
\begin{equation}
    \IrrBracket{\State}{\TotalEnergy} =\pdv{\State}{\State} \FrictionMatrix \pdv{\TotalEnergy}{\State} =\zeta_{\alpha\beta,\mu\nu} \pdv{\State}{x_\alpha} \pdv{\TotalEnergy}{x_\beta} \pdv{\TotalEnergy}{x_\mu} \pdv{\TotalEnergy}{x_\nu}=\zeta_{\alpha\beta,\mu\nu}  \pdv{\TotalEnergy}{x_\beta} \pdv{\TotalEnergy}{x_\mu} \pdv{\TotalEnergy}{x_\nu}=0.
\end{equation}

\subsection{Alternative parameterization -- penalty-based method}\label{sec:penalty}
An alternative strategy to incorporate GENERIC structure is to enforce the degeneracy condition by soft penalty as advocated in \cite{hernandez2021structure}. In this approach, $\TotalEnergy$, $\Entropy$, $\PoissonMatrix$, and $\FrictionMatrix$, may be approximated independently of each other. Again, $\TotalEnergy$ and $\Entropy$ are parameterized as neural networks ($\TotalEnergyNN$ and $\EntropyNN$), and $\PoissonMatrix$ and $\FrictionMatrix$ are parameterized as skew-symmetrizations/symmetrizations of matrices $\PoissonNNParam$ and $\FrictionNNParam$ with learnable entries as follows
\begin{equation}
\PoissonMatrixNN = \frac{1}{2}\left(\PoissonNNParam - \PoissonNNParam\Transpose \right) \quad \mathrm{and} \quad \FrictionMatrixNN = \FrictionNNParam \FrictionNNParam\Transpose.
\end{equation}
With this parameterization, the degeneracy conditions are simply enforced by minimizing two penalty terms, $\left\| \PoissonMatrixNN \pdv{\TotalEnergyNN}{\State} \right\|$ and  $\left\| \FrictionMatrixNN \pdv{\EntropyNN}{\State} \right\|$. We stress that this penalty will be enforced only to within optimization error.

If we write a system of neural ODEs as $\pdv{\State}{\TimeSymb} = \DynamicsNN$, where $\NNParams$ consists of learnable parameters, then Table 1 summarizes the components comprising $\DynamicsNN$ for black-box NODE, the penalty-base method, and GENERIC NODE (GNODE).
\begin{table}[h]
  \caption{Model summary}
  \label{tab:summary}
  \centering
  \begin{tabular}{cccc}
    \toprule
         & NODE & Penalty & GNODE\\
    \midrule
    $\DynamicsNN$ & $\DynamicsNN = \DynamicsNODE$ & $\DynamicsNN=\PoissonMatrixNN\pdv{\TotalEnergyNN}{\State} + \FrictionMatrixNN\pdv{\EntropyNN}{\State}$ & $\DynamicsNN= \PoissonBracket{\State}{\TotalEnergy} + \IrrBracket{\State}{\Entropy}$\\
    \midrule
    \multirow{2}{*}{Components} & \multirow{2}{*}{black-box MLP} & $\TotalEnergyNN$ and $\EntropyNN$ (MLPs) & $\TotalEnergyNN$ and $\EntropyNN$ (MLPs)\\
    & & $\PoissonMatrixNN$ and $\FrictionMatrixNN$ (2-tensor) & $\xi$ (3-tensor), $\Lambda$ and $D$ (2-tensor)\\
    \midrule
    $\NNParams$ &  $\NNParams = \NNParamsNODE$ & $\NNParams = \{\EnergyNNParam, \EntropyNNParam, \PoissonNNParam, \FrictionNNParam\}$ & $\NNParams =  \{\EnergyNNParam, \EntropyNNParam, \xi, \Lambda, D\}$\\
    \bottomrule
  \end{tabular}
\end{table}

\section{Experiments}
In this section, we assess the performance of the three parameterizations of the ODE dynamics which apply progressively more stringent priors. We implement the algorithms in \textsc{Python 3.6.5}, \textsc{NumPy 1.16.2}, and \textsc{PyTorch 1.7.1} \cite{paszke2017automatic}. For the time integrator, we use a \textsc{PyTorch} implementation of differentiable ODE solvers, \text{TorchDiffEq} \cite{chen2018neural}. All experiments are performed on \textsc{Macbook Pro} with 2.9 GHz i9 CPU and 32 GB memory. 

\subsection{Dataset and training}
The states $\State$ of GENERIC systems may generally be partitioned between ``observable'' states (e.g., position and momentum variables) denoted by $\Observable$ and ``non-observable'' states (e.g., entropy, configuration variables, etc) denoted by $\Unobservable$, i.e., $\State = [\Observable{}\Transpose, \Unobservable{}\Transpose]\Transpose$. We assume that training data is only available for the observable states, with the non-observable states functioning as hidden variables during training. For each benchmark problem, we take as manufactured training data a single trajectory of observable states obtained by integrating a reference ODE with known GENERIC structure from a known initial condition. We then split the sequence into three segments, $[0, \TrainTime]$, $(\TrainTime,\ValTime]$, and $(\ValTime,\TestTime]$ for training, validation, and test such that $0<\TrainTime<\ValTime<\TestTime$.

We employ mini-batching to train all three considered architectures. Each mini-batch consists of multiple short sequences of length $\batchlength$ whose initial conditions are randomly chosen from $[0, \TrainTime]$.  To train ``black-box'' neural ODEs, we simply use a stochastic gradient descent (SGD) optimizer to update the network weights and biases using the mini-batches on the observable states, $\{\Observable_{\ell},\Observable_{\ell+1},\ldots,\Observable_{\ell+\batchlength-1}\}$.

\begin{algorithm}[h]
\small
\SetAlgoLined
\caption{Neural ODE training}\label{alg:train1}
Initialize $\NNParams$\\
\For{$(i = 0;\ i < \maxepoch;\ i = i + 1)$}{
    Sample initial points $\{\Observable_{\ell(k)}\}_{k=1}^{\batchsize}$, where $\ell(k) \in [0, \TrainTime-\batchlength-1]$ for  $k=1,\ldots,\batchsize$\\
    $\ApproxObservable_{\ell(k)+1}$,\ldots,$\ApproxObservable_{\ell(k)+\batchlength}$ =  ODESolve($\Observable_{\ell(k)}$,$\DynamicsNN$,$\TimeSymb_1$,\ldots,$\TimeSymb_{\batchlength}$) for  $k=1,\ldots,\batchsize$\\
    Compute loss: $\LossSymb(\Observable_{\ell(k)+m}, \ApproxObservable_{\ell(k)+m})$\\ 
    Update $\NNParams$ via SGD
    }
\end{algorithm}

As opposed to the black-box neural ODEs, training the penalty-based approach and the GENERIC approach requires data to impose mini-batch initial conditions on non-observable states, i.e., $\{\State_{\ell},\State_{\ell+1},\ldots,\State_{\ell+\batchlength-1}\}$ with $\State_{\ell} = [\Observable_{\ell},\Unobservable_{\ell}]\Transpose$, where $\{\Unobservable_{\ell}\}$ are unavailable.  To address this issue, we propose a training strategy that alternately updates the model parameters and infers the non-observable states. We start with a guess for the non-observable states. We then alternate between (1) updating the model parameters using SGD while fixing the current non-observable states and (2) updating the non-observable states by solving an initial value problem using the most recent model. 

\begin{algorithm}[h]
\small
\SetAlgoLined
\caption{Penalty or GENERIC training}\label{alg:train2}
Initialize $\NNParams$ and $\{\Unobservable_0,\ldots,\Unobservable_{\TrainTime}\}$\\
Construct a dataset as $\State_i = [\Observable_i{}\Transpose, \Unobservable_i{}\Transpose]\Transpose$, for $i=0,\ldots,\TrainTime$\\ 
\For{$(i = 0;\ i < \maxepoch;\ i = i + 1)$}{
    Sample initial points $\{\State_{\ell(k)}\}_{k=1}^{\batchsize}$, where $\ell(k) \in [0, \TrainTime-\batchlength-1]$ for  $k=1,\ldots,\batchsize$\\
    $\ApproxState_{\ell(k)+1}$,\ldots,$\ApproxState_{\ell(k)+\batchlength}$ =  ODESolve($\State_{\ell(k)}$,$\DynamicsNN$,$\TimeSymb_1$,\ldots,$\TimeSymb_{\batchlength}$) for  $k=1,\ldots,\batchsize$\\
    Compute loss: $\LossSymb(\Observable_{\ell(k)+m}, \ApproxObservable_{\ell(k)+m})$\\  
    Update $\NNParams$ via SGD\\
    \If{$i \bmod \nupdate == 0$} {
    $\ApproxState_{1}$,\ldots,$\ApproxState_{\TrainTime}$ =  ODESolve($\State_{0}$,$\DynamicsNN$,$\TimeSymb_1$,\ldots,$\TrainTime$)\\
    Update a dataset as $\State_i = [\Observable_i{}\Transpose, \ApproxUnobservable_i{}\Transpose]\Transpose$, for $i=0,\ldots,\TrainTime$
    }
}
\end{algorithm}

For \text{ODESolve}, we use the Dormand--Prince method (dopri5) \cite{dormand1980family} with relative tolerance $10^{-5}$ and absolute tolerance $10^{-6}$. The loss function $\LossSymb$ measures the discrepancy between the ground truth states and approximate states via mean absolute errors, and the network weights and biases are updated using Adamax  \cite{kingma2014adam} with an initial learning rate 0.01.

In the following, we test the proposed algorithms with two benchmark problems: a damped nonlinear oscillator and two gas containers problems. Data for all considered benchmark problems can be found in  \cite{shang2020structure}.

\subsection{Damped nonlinear oscillator}\label{sec:dno}
As a first benchmark problem, we consider a damped nonlinear oscillator which exhibits a natural GENERIC structure: 
\begin{equation}\label{dnoEqn}
        \dv{\PositionSymb}{\TimeSymb} = \frac{\MomentumSymb}{m}, \qquad 
        \dv{\MomentumSymb}{\TimeSymb} = k\sin(\PositionSymb) - \gamma \MomentumSymb, \qquad \dv{\Entropy}{\TimeSymb} = \frac{\gamma\PositionSymb^2}{mT},
\end{equation}
where $(\PositionSymb,\MomentumSymb)$ denote the position and momentum of the particle, and $\Entropy$ is the entropy of the surrounding thermal bath. The constant parameters $m$, $\gamma$, and $T$ represent the mass of the particle, the damping rate, and the constant temperature of the thermal bath. The total energy of the GENERIC system is 
\begin{equation}
    \TotalEnergy(\PositionSymb,\MomentumSymb,\Entropy) =  \Hamiltonian(\PositionSymb,\MomentumSymb) + T\Entropy = \frac{\MomentumSymb^2}{2m} - k \cos(\PositionSymb) + T\Entropy,
\end{equation}
where $\Hamiltonian(\PositionSymb,\MomentumSymb)$ is the Hamiltonian of the particle (the sum of the kinetic and the potential energy). 

\begin{figure}[!t]
    \centering
    \subfigure[Trajectory]{\includegraphics[scale=.55]{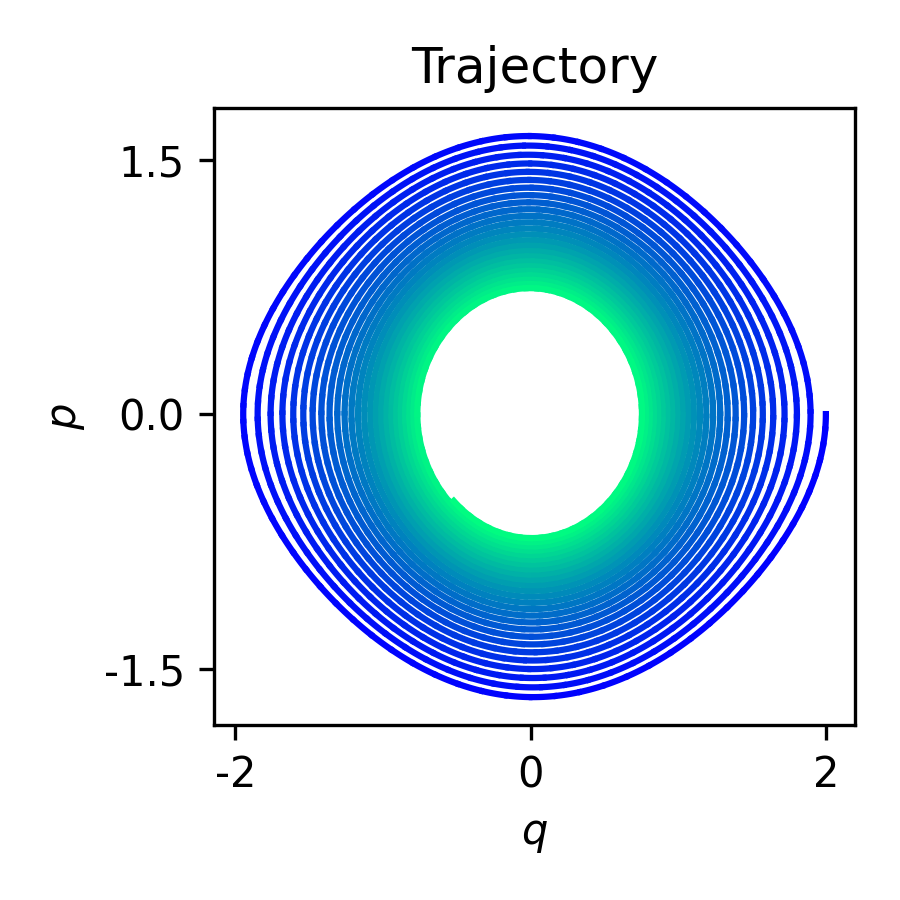}\label{fig:dno_traj}}
    \subfigure[$\dv{\Entropy}{\TimeSymb}$ -- penalty]{\includegraphics[scale=.55]{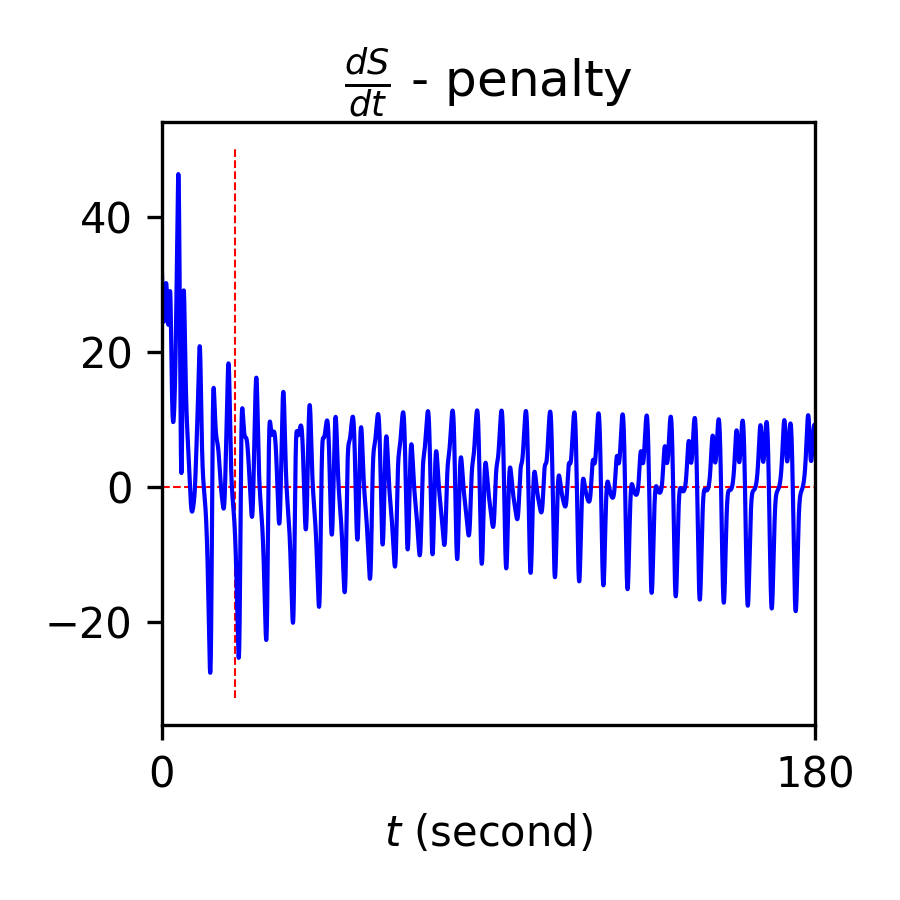}}
    \subfigure[$\dv{\Entropy}{\TimeSymb}$ -- GNODE]{\includegraphics[scale=.55]{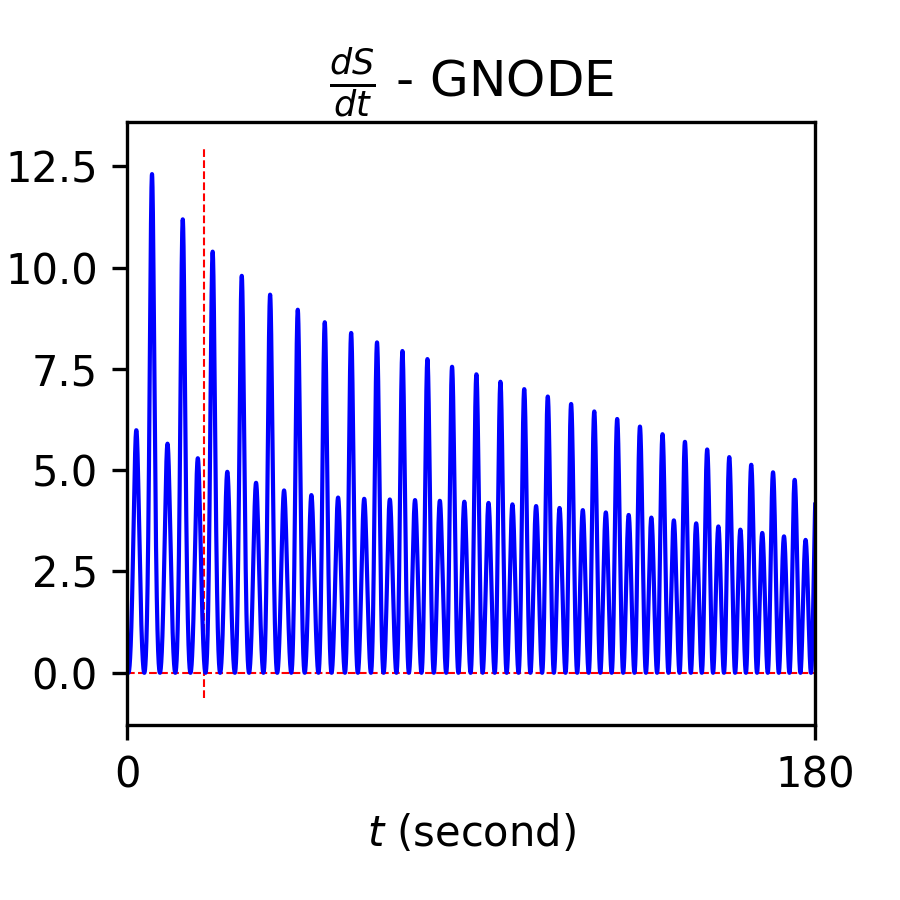}}\\
    \subfigure[$\Hamiltonian$ -- NODE]{\includegraphics[scale=.525]{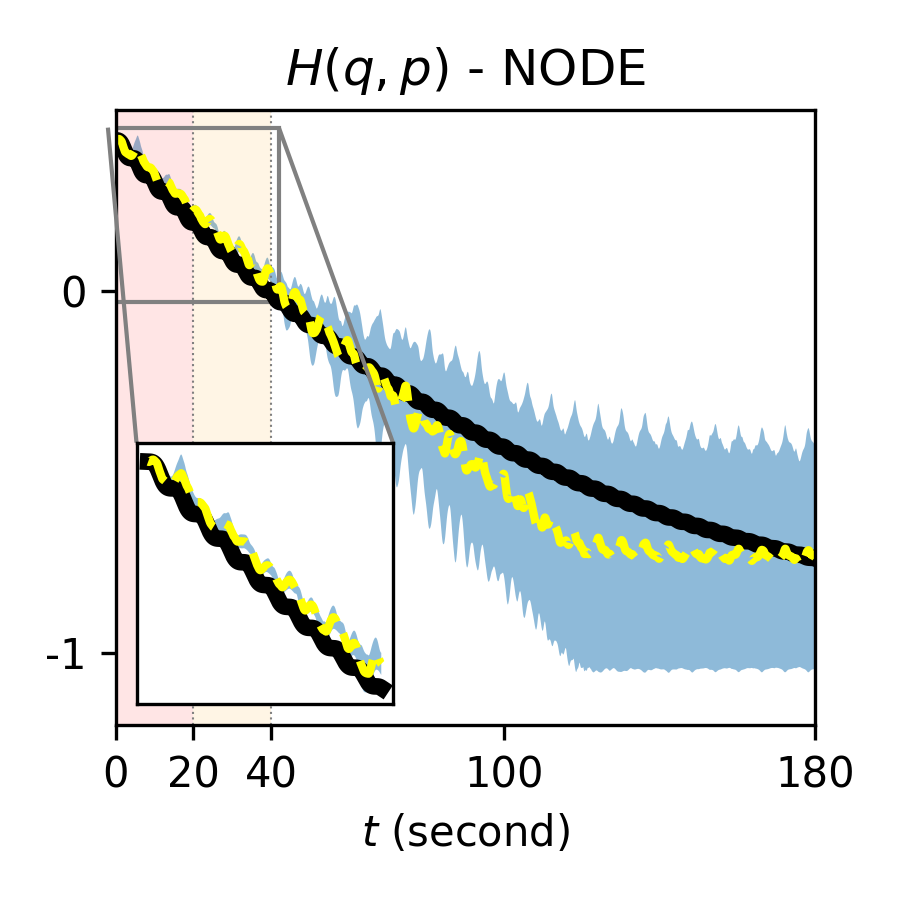}}
    \subfigure[$\Hamiltonian$ -- penalty]{\includegraphics[scale=.525]{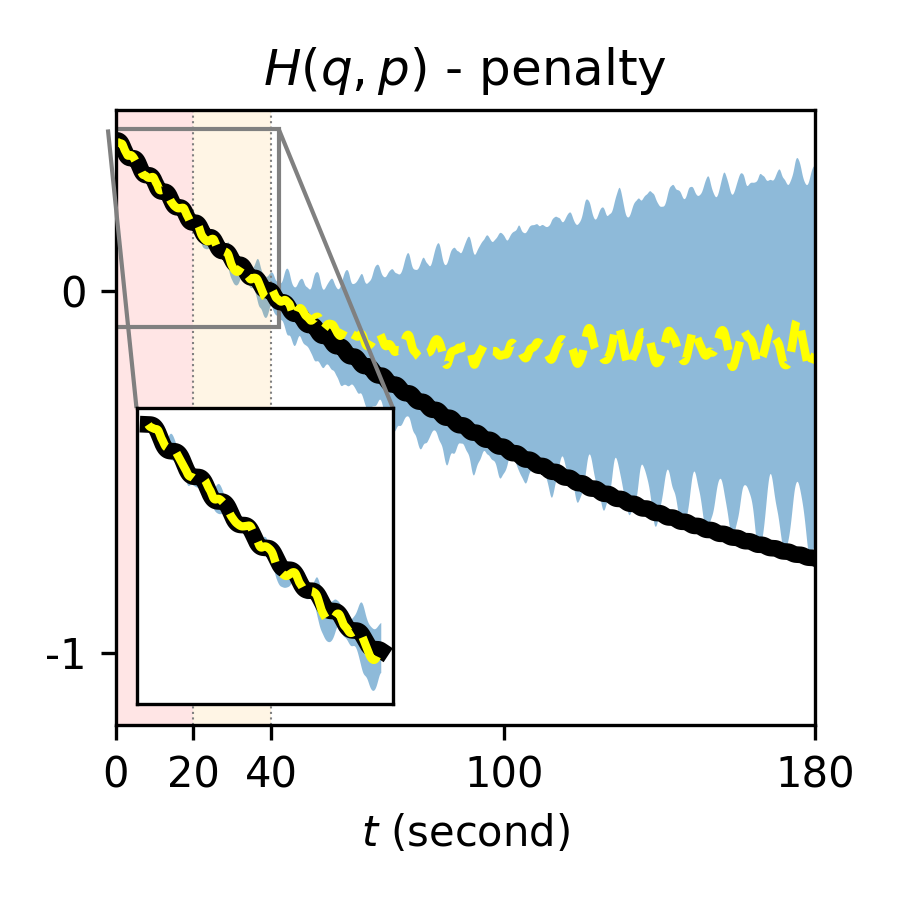}}
    \subfigure[$\Hamiltonian$ -- GNODE]{\includegraphics[scale=.525]{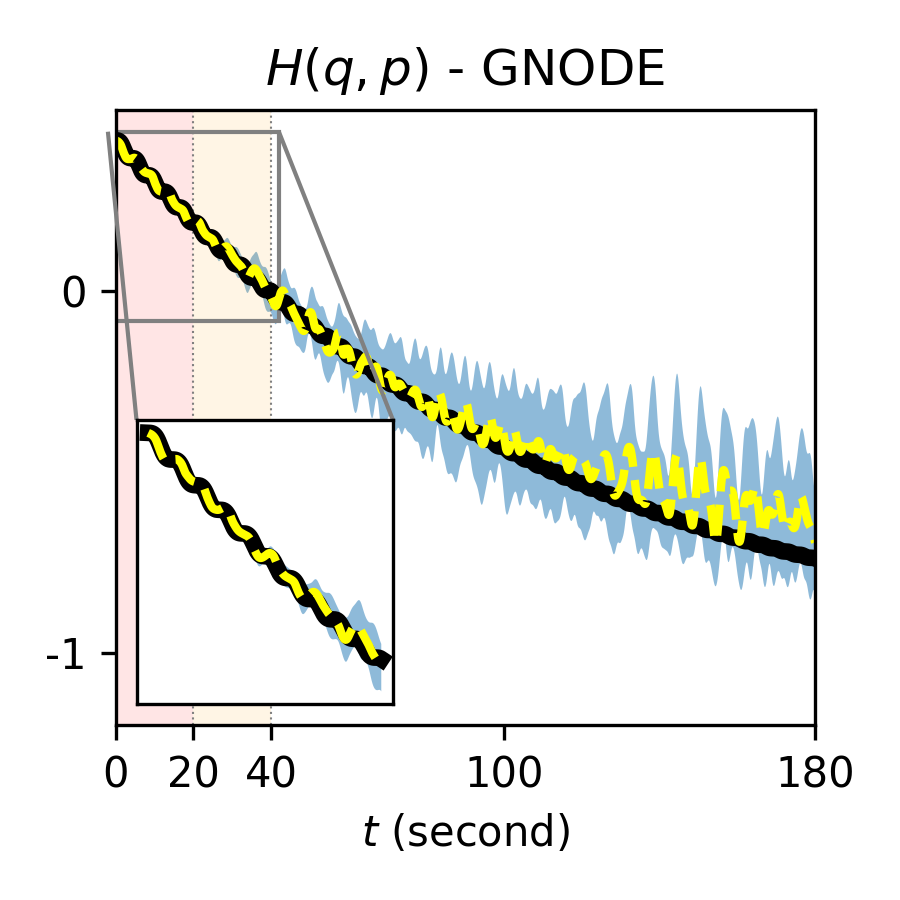}}\\
    \includegraphics[scale=.75]{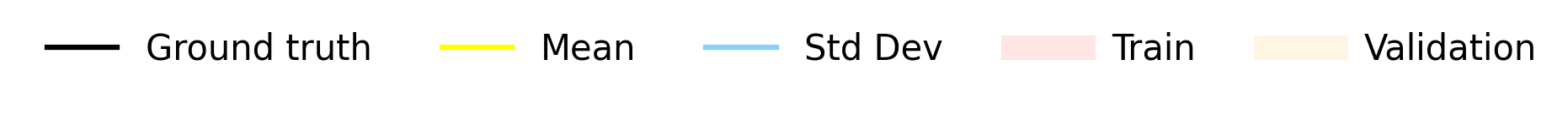}
    \caption{For the damped nonlinear oscillator, the physical entropy may be evaluated via the formula $E=H+TS$. While all three methods fit training data reasonably well, NODE and the penalty approach rapidly deviate. An inspection of $\frac{dS}{dt}$ for the penalty method shows that the soft penalty is insufficient to ensure compatibility with the second law. GNODE is able to consistently learn an entropy $S$ which closely tracks the physical entropy $(E-H)/T$.}
    \label{fig:dno}
\end{figure}

In this benchmark problem, the observable states consist of the position and the momentum variables, i.e., $\Observable=[\PositionSymb, \MomentumSymb]\Transpose$. We consider a single non-observable variable, i.e., $\Unobservable = s$. Now, our goal is to learn a system of ODEs that conforms the GENERIC structure described in Section \ref{sec:param_generic} and infer the non-observable variable via Algorithm \ref{alg:train2}. That is, for GNODE, we model  $\TotalEnergyNN$ and $\EntropyNN$ to take $\State = [\PositionSymb, \MomentumSymb,s]\Transpose$ as an input.

For black-box NODEs, we consider an MLP with 4 hidden layers with 5 neurons in each layer and hyperbolic tangent (Tanh) activation function. For the penalty-based approach, we consider MLPs with 3 hidden layers with 5 neurons in each layer and Tanh for parameterizing $\TotalEnergyNN$ and $\EntropyNN$, and $3\times 3$ learnable entries for $\PoissonMatrixNN$ and $\FrictionMatrixNN$. We add the penalty terms (see Section \ref{sec:penalty}) that are weighted by $10^{-4}$ to the main loss objective. Lastly, for the GENERIC approach, we consider an MLP with 1 hidden layer with 10 neurons and Tanh for parameterizing $\TotalEnergyNN$, and a linear layer for parameterizing $\EntropyNN$. Then, we use $3\times 3 \times 3$ skew-symmetric tensor to parameterize $\xi$, $3\times 3$ skew-symmetric tensor to parameterize $\Lambda$, and $3\times  1$ tensor, $d$, to parameterize $D$, i.e., $D = dd\Transpose$. For initializing layers in MLPs, we use the \textsc{PyTorch} default uniform distribution and, for initializing learnable entries, we initialize them with unit normal distribution. We initialize the non-observable variable as $\Unobservable_\ell = s_\ell = t_\ell$ (i.e., setting it to be monotonically increasing) in Line 1 of Algorithm~\ref{alg:train2}.

The dataset consists of a sequence of 180,000 timesteps with $\FinalTime=180$ (in second) and step size $\TimeStep=0.001$. We then split the dataset into training, validation, and testing sets such that $\TrainTime=20$, $\ValTime=40$, and $\TestTime=180$. Each mini-batch consists of $\batchsize=20$ subsequences of length $\batchlength=120$. The maximum training step is set as $\maxepoch=30000$ and the update is performed at every $\nupdate=500$ training steps (in Algorithm \ref{alg:train2}).

In the experiment, we consider $m=k=T=1$, and $\gamma=0.01$. The initial condition is given as $\InitState=[2,0,0]\Transpose$, where the initial condition for the non-observable variable is arbitrarily set. Results of the comparison are given in Figure \ref{fig:dno}. The results are obtained from 5 independent runs (i.e., 5 different random seeds). The plots of $\frac{dS}{dt}$ reveal that the soft enforcement of constraints in the penalty formulation leads to negative entropy production of large magnitude, while the GNODE approach enforces by construction $\frac{dS}{dt}\geq 0$. When extrapolating well beyond the training time interval, both black-box NODE and the penalty approach have a large standard deviation in the predicted $H$. GNODE in contrast learns a nearby entropy which consistently dissipates the correct amount of energy.

\begin{figure}[!t]
    \centering
    \subfigure[Trajectory]{\includegraphics[scale=.55]{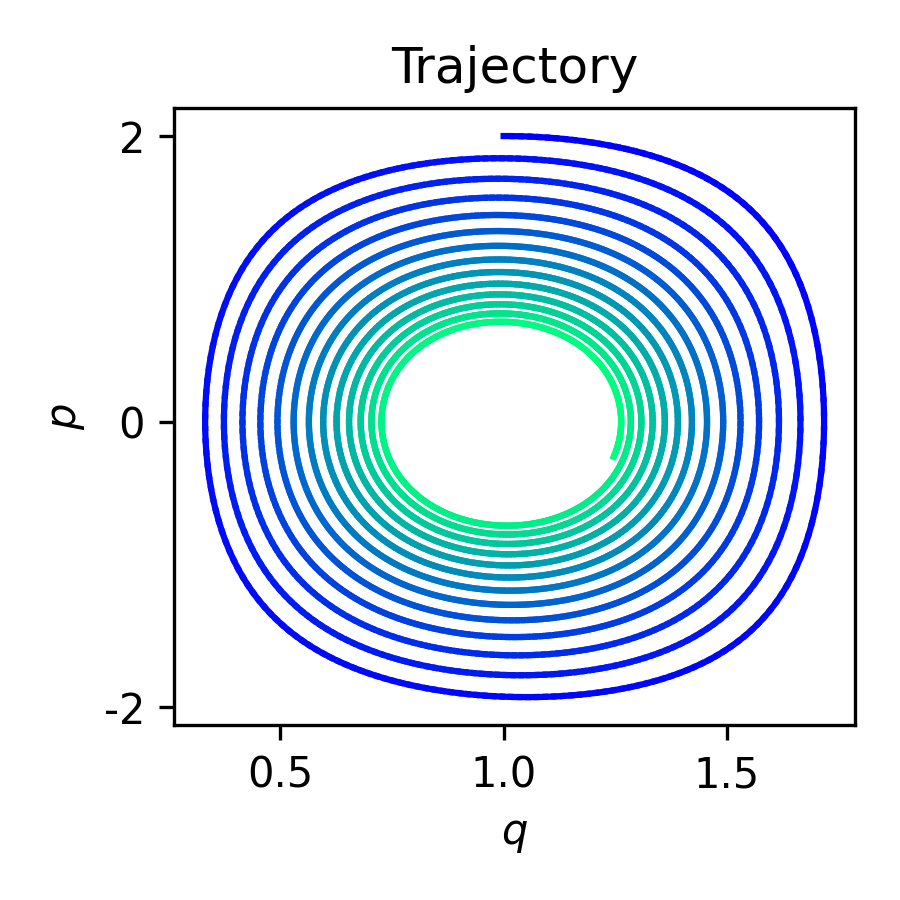}\label{fig:tgc_traj}}
    \subfigure[$\dv{\Entropy}{\TimeSymb}$ -- penalty]{\includegraphics[scale=.55]{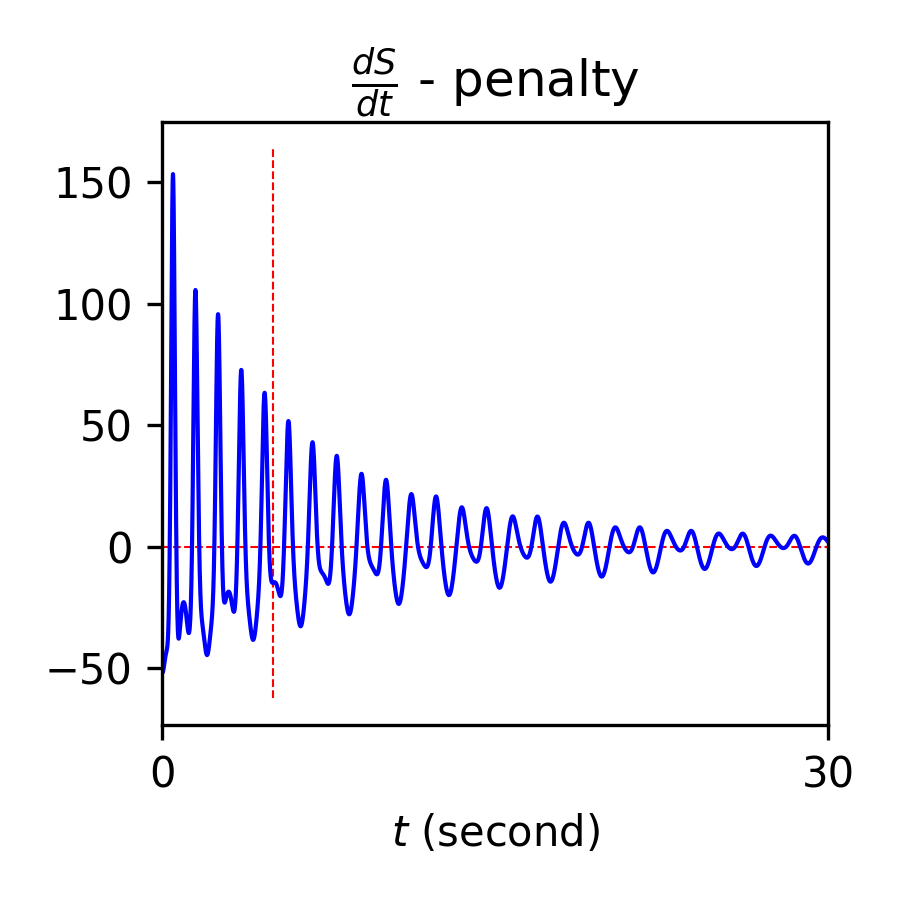}}
    \subfigure[$\dv{\Entropy}{\TimeSymb}$ -- GNODE]{\includegraphics[scale=.55]{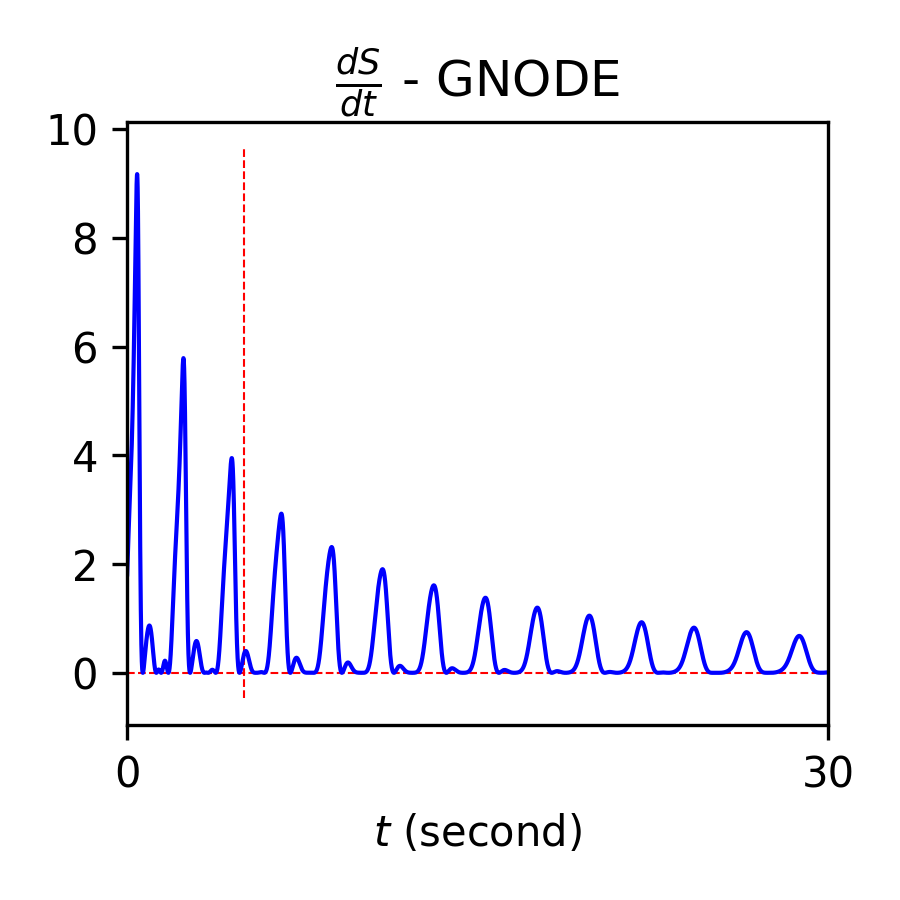}}
    \\
    \subfigure[$\Hamiltonian$ -- NODE]{\includegraphics[scale=.55]{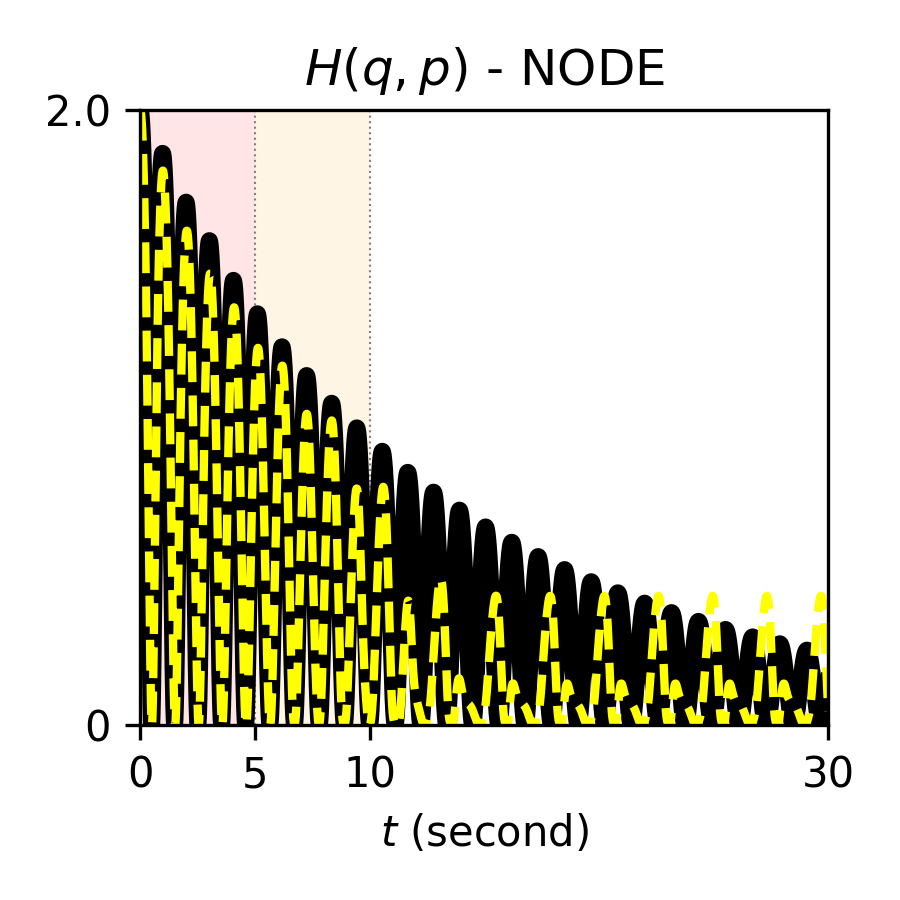}}
    \subfigure[$\Hamiltonian$ -- penalty]{\includegraphics[scale=.55]{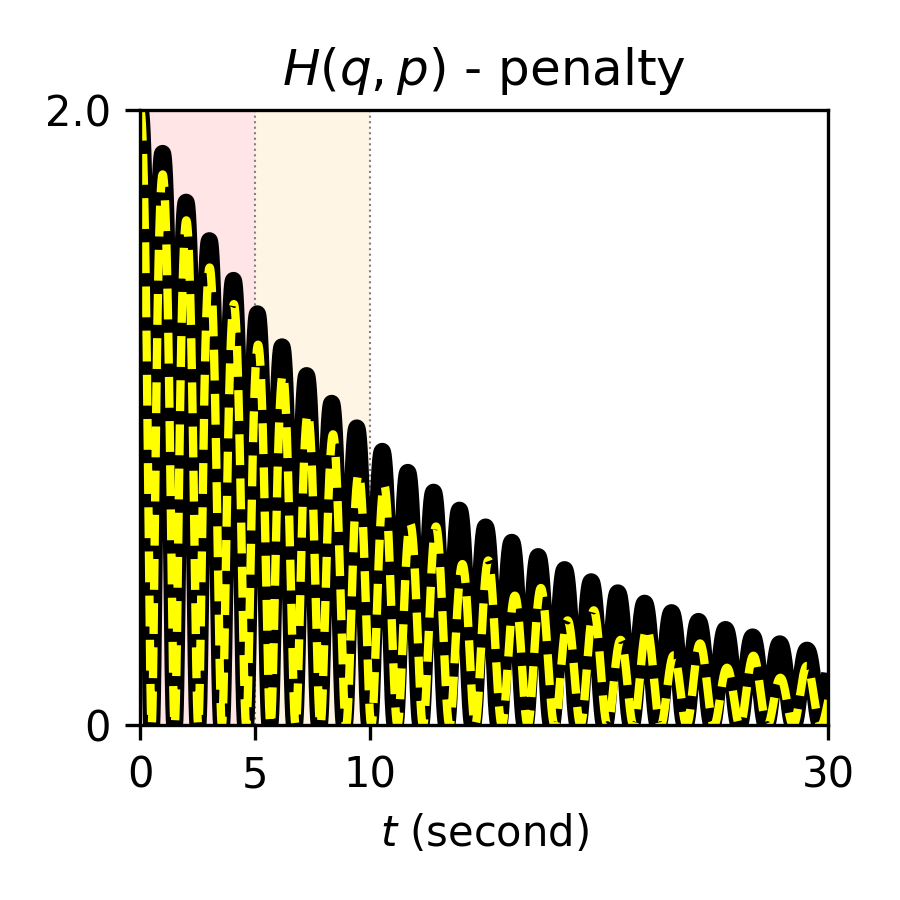}}
    \subfigure[$\Hamiltonian$ -- GNODE]{\includegraphics[scale=.55]{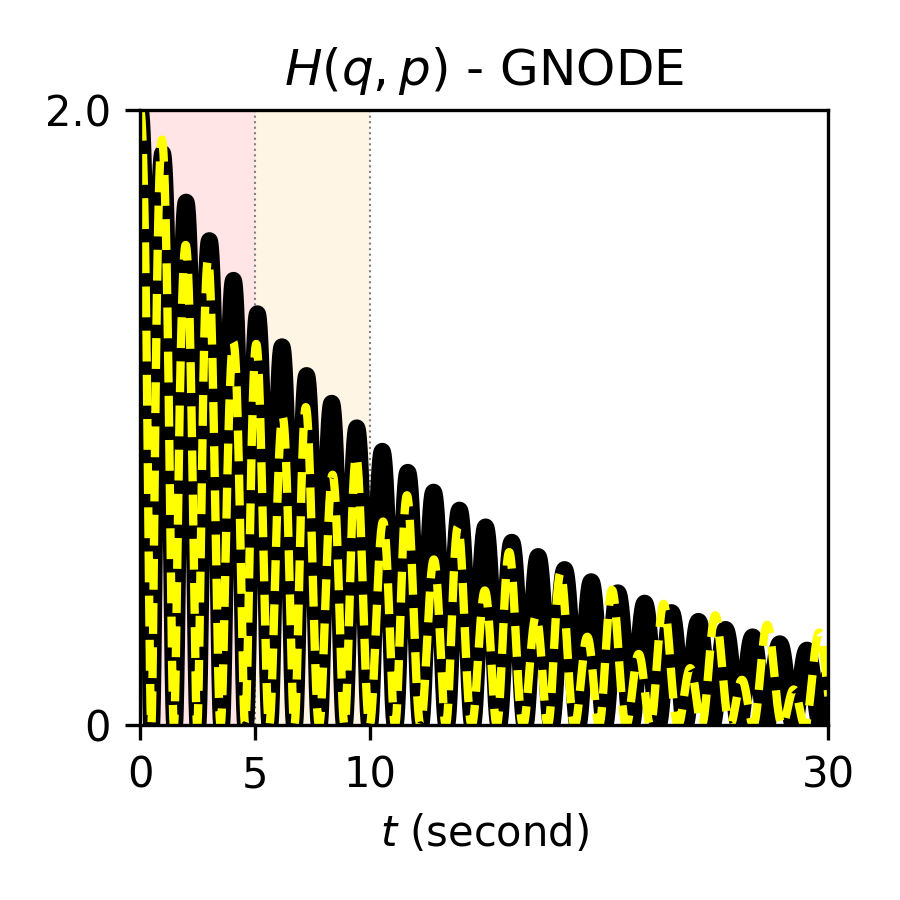}}\\
    \includegraphics[scale=.75]{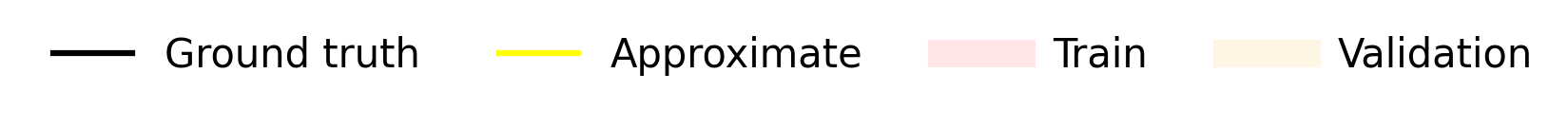}
    \caption{For the two gas cylinder problem the network must learn a significantly more complex entropy-energy relationship than in the damped oscillator. The results however are similar; GNODE provides remarkable improvement in forecasting due to its faithful reproduction of the second law of thermodynamics \textit{(top row, center + right)}.} 
    \label{fig:tgc}
\end{figure}

\subsection{Two gas containers}
The second benchmark problem considers two (ideal) gas containers, separated by a moving wall, exchanging heat and volume. Here, we are interested in the position and the momentum of the separating wall, i.e., $\Observable = [\PositionSymb,\MomentumSymb]\Transpose$. This problem possesses a highly nonlinear expression for the entropy \cite{shang2020structure}:
\begin{alignat}{3}
        \dv{\PositionSymb}{\TimeSymb} &= \frac{\MomentumSymb}{m}, \qquad &&\dv{\MomentumSymb}{\TimeSymb} &&= \frac{2}{3}\left(\frac{\TotalEnergy_1}{\MomentumSymb} - \frac{\TotalEnergy_2}{2L_{g}-\MomentumSymb} \right),\\
        \dv{\Entropy_1}{\TimeSymb} &= \frac{9N^2k_B^2\alpha}{4\TotalEnergy_1} \left(\frac{1}{\TotalEnergy_1} -\frac{1}{\TotalEnergy_2}  \right),  \qquad &&\dv{\Entropy_2}{\TimeSymb} &&= - \frac{9N^2k_B^2\alpha}{4\TotalEnergy_1} \left(\frac{1}{\TotalEnergy_1} -\frac{1}{\TotalEnergy_2}  \right), 
\end{alignat}
where $(\PositionSymb,\MomentumSymb)$ denote the position and momentum of the separating wall and $\Entropy_1$ and $\Entropy_2$ are the entropies of the two subsystems. The constants $m$ denotes the mass of the wall, 2$L_g$ is the total length of the two containers. Following \cite{shang2020structure}, we set $N k_B=1$, which fixes a characteristic macroscopic unit of entropy, and $\alpha=0.5$. The internal energies of the two subsystems has the relationship with the associated entropies and volumes via the Sackur--Tetrode equation for ideal gases such that 
\begin{equation}
    \frac{\Entropy_i}{Nk_B} = \ln \left[\hat cV_i(E_i)^{3/2} \right], \quad i=1,2,
\end{equation}
where $\hat c$ is a constant to ensure the argument of the logarithm dimensionless (set as $\hat c = 102.25$). The total energy is given by 
\begin{equation}
    \TotalEnergy(\PositionSymb,\MomentumSymb,\Entropy_1,\Entropy_2) = \Hamiltonian(\PositionSymb,\MomentumSymb) + \TotalEnergy_1 + \TotalEnergy_2= \frac{\MomentumSymb^2}{2m} + \TotalEnergy_1 + \TotalEnergy_2.
\end{equation}

The observable states consist of the position and the momentum variables, i.e., $\Observable=[\PositionSymb, \MomentumSymb]\Transpose$. We consider two non-observable variables, i.e., $\Unobservable = [s_1, s_2]\Transpose$. This problem is more challenging as the dynamics of the observable variables strongly depends on the dynamics of the non-obesrvable variables. Now we train GNODE to learn a system of ODEs that conforms the GENERIC structure described in Section \ref{sec:param_generic} and infer the non-observed variables via Algorithm \ref{alg:train2}. Again, for GNODE, we model  $\TotalEnergyNN$ and $\EntropyNN$ to take $\State = [\PositionSymb, \MomentumSymb,s_1,s_2]\Transpose$ as an input.

For black-box NODEs, we consider an MLP with 4 hidden layers with 5 neurons in each layer and Tanh as nonlinearity. For the penalty-based approach, we consider MLPs with 3 hidden layers with 5 neurons in each layer and Tanh for parameterizing $\TotalEnergyNN$ and $\EntropyNN$, and $3\times 3$ learnable entries for $\PoissonMatrixNN$ and $\FrictionMatrixNN$. We add the penalty terms that are weighted by $1e-4$ to the main loss objective. Lastly, for the GENERIC approach, we consider an MLP with 2 hidden layer with 5 neurons and Tanh for parameterizing $\TotalEnergyNN$, and 1 hidden layer with 5 neurons and Tanh for parameterizing $\EntropyNN$. Then, we use $3\times 3 \times 3$ skew-symmetric tensor to parameterize $\xi$, $4\times 4$ skew-symmetric tensor to parameterize $\Lambda$, and $4\times  1$ tensor, $d$, to parameterize $D$, i.e., $D = dd\Transpose$. For initializing layers in MLPs, we use the \textsc{PyTorch} default uniform distribution and, for initializing learnable entries, we initialize them with unit normal distribution. The non-observable variables are initialized as $s_{1,\ell} = s_{2,\ell} = t_\ell$.

The dataset consists of a sequence of 30,000 timesteps with $\FinalTime=30$ seconds and step size $\TimeStep=0.001$. We then split the dataset into training, validation, and testing subsets such that $\TrainTime=5$, $\ValTime=10$, and $\TestTime=30$. Each mini-batch consists of $\batchsize=20$ subsequences of length $\batchlength=40$. The maximum training step is set as $\maxepoch=50000$ and the update is performed at every $\nupdate=500$ training steps (in Algorithm \ref{alg:train2}).

We set $m=L_g=1$ and the initial condition is given as $\InitState=[1,2,0,0]$. Again, the initial condition for the non-observable variables are set arbitrarily. The results are depicted in Figure \ref{fig:tgc} and demonstrate similar results to the damped oscillator. We depict the results from the best performing instances for each model out of five independent runs. Both NODE and the penalty method provide inaccurate forecasting beyond the training set, and the penalty method can be seen to generate negative entropy violating the second law of thermodynamics, while $\frac{dS}{dt}\geq 0$ holds strictly for GNODE.

\subsection{Stochastic damped harmonic oscillator}

We finally fit a GNODE model to the system considered Section \ref{sec:dno} and use the learned $E$, $S$, $M$ and $L$ in the right hand side of the SDE in \eqref{eq:genericSDE}, and compare as a baseline to using instead the analytical $E$, $S$, $M$, and $L$ from \eqref{dnoEqn}. This amounts to driving both the true system and data-driven dynamics with thermal noise which exactly balances the dissipation, and requires the FDT to hold to realize stationary statistics.

\begin{figure}[!t]
    \centering
    \subfigure[$q$ -- analytic ]{\includegraphics[scale=.45]{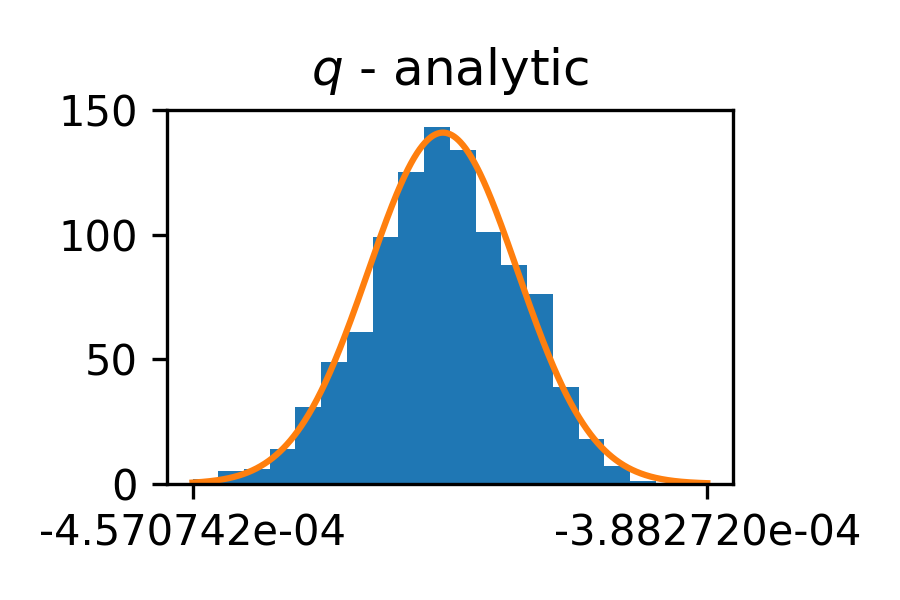}}
    \subfigure[$q$ -- GNODE ]{\includegraphics[scale=.45]{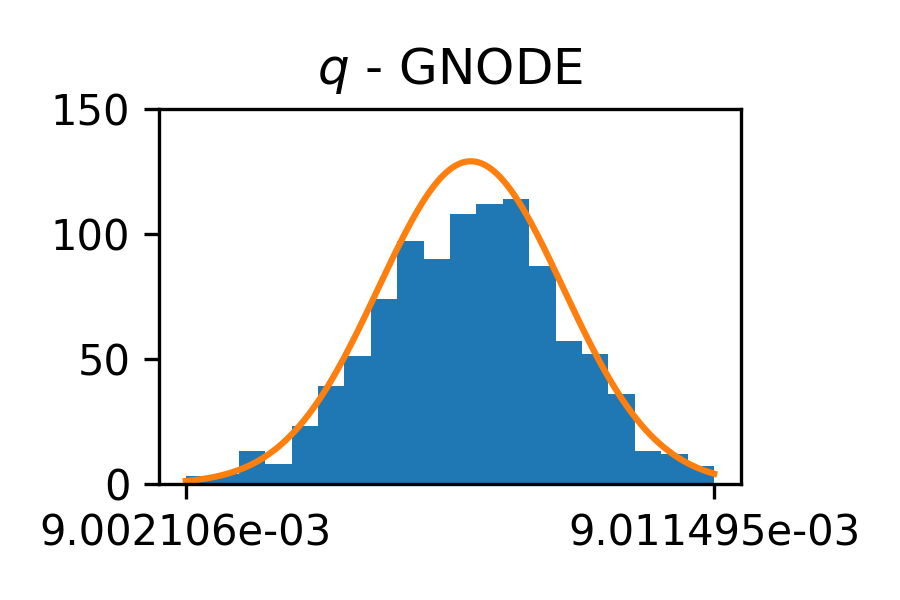}}
    \subfigure[$p$ -- analytic ]{\includegraphics[scale=.45]{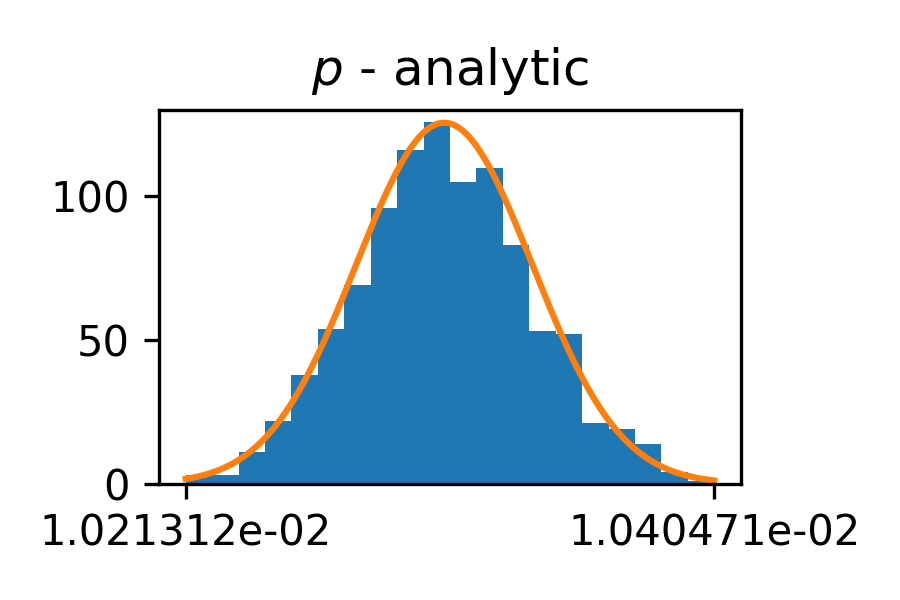}}
    \subfigure[$p$ -- GNODE ]{\includegraphics[scale=.45]{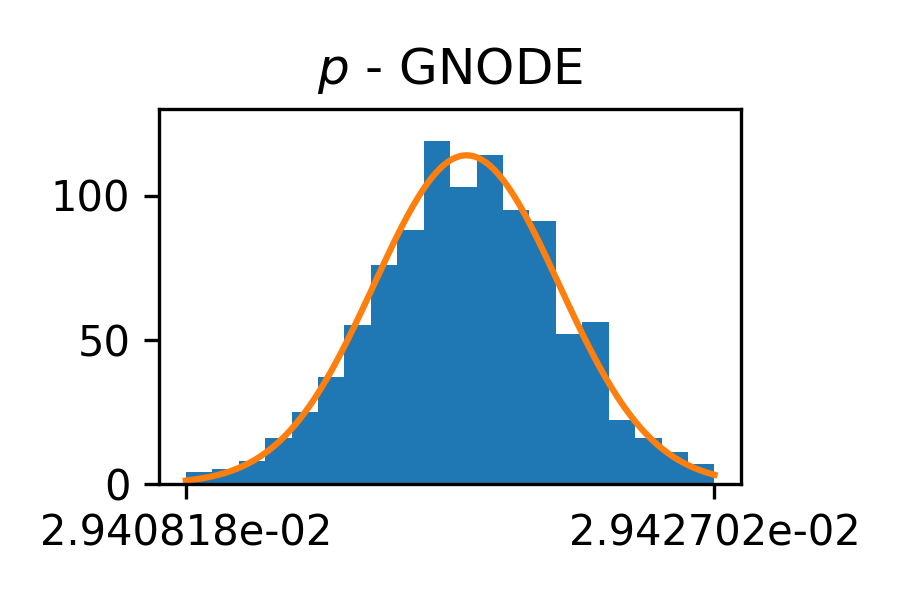}}
    \caption{Distribution of $q$ and $p$ at $\FinalTime=80$ for the SDE \eqref{eq:genericSDE}. The standard deviations of $(q,p)$ for baseline and GNODE are ($9.9684\times 10^{-6}$,\,$3.1971\times 10^{-5}$) and ($1.6436\times 10^{-6}$,\,$3.2972\times10^{-6}$), respectively.} 
    \label{fig:sde}
\end{figure}

In this experiment, we consider the damped nonlinear oscillator with $m=10$, and $\gamma=0.16$ and the same neural network architecture considered in Section \ref{sec:dno}. We use a sequence of 80,000 timesteps ($\FinalTime=80$ and $\TimeStep=0.0001$) to train the neural network. We use the same training strategy that is used in Section \ref{sec:dno} (Algorithm \ref{alg:train2}); the only difference is that we use the mini-batch of size $\batchsize=40$.

For Figure \ref{fig:sde} we obtain statistics from solving \eqref{eq:genericSDE} with $\FinalTime=80$ of Euler--Maruyama \cite{kloeden1992stochastic} with step size $\TimeStep=0.001$. The mean of the resulting SDE solutions show substantial deviation, consistent with the fact that training is performed only on deterministic training data and not the SDE, which is compounded by nonlinearities in the data. However, the standard deviation of the distribution shows good agreement, suggesting that the FDT enforces thermodynamically consistent energy budget between dissipation and stochastic forcing.

\section{Conclusions}

We have constructed a generalization of structure preserving networks for reversible dynamics to handle dissipative processes. Unlike the reversible case, the bracket structure requires a much more careful treatment of degeneracy conditions to ensure compatibility with the first and second laws of thermodynamics. Numerical examples show that our novel parameterization is able to provide non-decreasing entropy that translates to improved robustness for out-of-distribution forecasting. We additionally show that exact treatment of dissipative processes allows introduction of thermal forcing which satisfies a discrete FDT. To achieve thermodynamically consistent equilibrium distributions in this setting, we have shown the degeneracy condition must be imposed exactly.

While this work establishes the value of imposing bracket structure for dissipative processes in terms of generalization, robustness, and physical realizability, the training approach applied here is applicable only to relatively small systems, restricting its application e.g. to learning dynamics of reduced-order models. Future work will focus on developing more scalable training strategies for learning ODEs of many variables.

\section{Acknowledgements}
Sandia National Laboratories is a multimission laboratory managed and operated by National Technology and Engineering Solutions of Sandia, LLC, a wholly owned subsidiary of Honeywell International, Inc., for the U.S. Department of Energy’s National Nuclear Security Administration under contract {DE-NA0003530}.  This paper describes objective technical results and analysis.  Any subjective views or opinions that might be expressed in the paper do not necessarily represent the views of the U.S. Department of Energy or the United States Government. 

We thank Chris Eldred for his recommendation to consider Ottingers bracket formulation of GENERIC. The work of N. Trask, and P. Stinis is supported by the U.S. Department of Energy, Office of Advanced Scientific Computing Research under the Collaboratory on Mathematics and Physics-Informed Learning Machines for Multiscale and Multiphysics Problems (PhILMs) project. N. Trask and K. Lee are supported by the Department of Energy early career program. 
{\small
\bibliographystyle{unsrt}
\bibliography{ref}

\begin{thebibliography}{10}

\bibitem{baker2019workshop}
Nathan Baker, Frank Alexander, Timo Bremer, Aric Hagberg, Yannis Kevrekidis,
  Habib Najm, Manish Parashar, Abani Patra, James Sethian, Stefan Wild, et~al.
\newblock Workshop report on basic research needs for scientific machine
  learning: Core technologies for artificial intelligence.
\newblock Technical report, USDOE Office of Science (SC), Washington, DC
  (United States), 2019.

\bibitem{rackauckas2020universal}
Christopher Rackauckas, Yingbo Ma, Julius Martensen, Collin Warner, Kirill
  Zubov, Rohit Supekar, Dominic Skinner, Ali Ramadhan, and Alan Edelman.
\newblock Universal differential equations for scientific machine learning.
\newblock {\em arXiv preprint arXiv:2001.04385}, 2020.

\bibitem{chen2018neural}
Ricky~TQ Chen, Yulia Rubanova, Jesse Bettencourt, and David Duvenaud.
\newblock Neural ordinary differential equations.
\newblock In {\em Proceedings of the 32nd International Conference on Neural
  Information Processing Systems}, pages 6572--6583, 2018.

\bibitem{Greydanus2019hnn}
Samuel Greydanus, Misko Dzamba, and Jason Yosinski.
\newblock Hamiltonian neural networks.
\newblock In {\em Advances in Neural Information Processing Systems},
  volume~32. Curran Associates, Inc., 2019.

\bibitem{cranmer2020lagrangian}
Miles Cranmer, Sam Greydanus, Stephan Hoyer, Peter Battaglia, David Spergel,
  and Shirley Ho.
\newblock Lagrangian neural networks.
\newblock In {\em ICLR 2020 Workshop on Integration of Deep Neural Models and
  Differential Equations}, 2020.

\bibitem{chen2019symplectic}
Zhengdao Chen, Jianyu Zhang, Martin Arjovsky, and L{\'e}on Bottou.
\newblock Symplectic recurrent neural networks.
\newblock In {\em International Conference on Learning Representations}, 2019.

\bibitem{jin2020sympnets}
Pengzhan Jin, Zhen Zhang, Aiqing Zhu, Yifa Tang, and George~Em Karniadakis.
\newblock Sympnets: Intrinsic structure-preserving symplectic networks for
  identifying hamiltonian systems.
\newblock {\em Neural Networks}, 132:166--179, 2020.

\bibitem{tong2021symplectic}
Yunjin Tong, Shiying Xiong, Xingzhe He, Guanghan Pan, and Bo~Zhu.
\newblock Symplectic neural networks in taylor series form for hamiltonian
  systems.
\newblock {\em Journal of Computational Physics}, page 110325, 2021.

\bibitem{haber2017stable}
Eldad Haber and Lars Ruthotto.
\newblock Stable architectures for deep neural networks.
\newblock {\em Inverse Problems}, 34(1):014004, 2017.

\bibitem{sanchez2019hamiltonian}
Alvaro Sanchez-Gonzalez, Victor Bapst, Kyle Cranmer, and Peter Battaglia.
\newblock Hamiltonian graph networks with ode integrators.
\newblock {\em arXiv preprint arXiv:1909.12790}, 2019.

\bibitem{guha2007metriplectic}
Partha Guha.
\newblock Metriplectic structure, leibniz dynamics and dissipative systems.
\newblock {\em Journal of Mathematical Analysis and Applications},
  326(1):121--136, 2007.

\bibitem{grmela1997dynamics}
Miroslav Grmela and Hans~Christian {\"O}ttinger.
\newblock Dynamics and thermodynamics of complex fluids. i. development of a
  general formalism.
\newblock {\em Physical Review E}, 56(6):6620, 1997.

\bibitem{ottinger2020framework}
Hans~Christian {\"O}ttinger, Mark~A Peletier, and Alberto Montefusco.
\newblock A framework of nonequilibrium statistical mechanics. i. role and
  types of fluctuations.
\newblock {\em Journal of Non-Equilibrium Thermodynamics}, 1(ahead-of-print),
  2020.

\bibitem{montefusco2018coarse}
Alberto Montefusco, Mark~A Peletier, and Hans~Christian {\"O}ttinger.
\newblock Coarse-graining via the fluctuation-dissipation theorem and
  large-deviation theory.
\newblock {\em arXiv preprint arXiv:1809.07253}, 2018.

\bibitem{ottinger2018generic}
Hans~Christian {\"O}ttinger.
\newblock Generic: Review of successful applications and a challenge for the
  future.
\newblock {\em arXiv preprint arXiv:1810.08470}, 2018.

\bibitem{lorenz1963deterministic}
Edward~N Lorenz.
\newblock Deterministic nonperiodic flow.
\newblock {\em Journal of atmospheric sciences}, 20(2):130--141, 1963.

\bibitem{grassberger1983characterization}
Peter Grassberger and Itamar Procaccia.
\newblock Characterization of strange attractors.
\newblock {\em Physical review letters}, 50(5):346, 1983.

\bibitem{gao2021non}
Zhen Gao, Qi~Liu, Jan~S Hesthaven, Bao-Shan Wang, Wai~Sun Don, and Xiao Wen.
\newblock Non-intrusive reduced order modeling of convection dominated flows
  using artificial neural networks with application to rayleigh-taylor
  instability.
\newblock 2021.

\bibitem{hesthaven2018structure}
Jan~S Hesthaven and Cecilia Pagliantini.
\newblock Structure-preserving reduced basis methods for hamiltonian systems
  with a nonlinear poisson structure.
\newblock {\em EPFL Infoscience}, 2018.

\bibitem{chorin2007problem}
Alexandre Chorin and Panagiotis Stinis.
\newblock Problem reduction, renormalization, and memory.
\newblock {\em Communications in Applied Mathematics and Computational
  Science}, 1(1):1--27, 2007.

\bibitem{lutter2019deep}
Michael Lutter, Christian Ritter, and Jan Peters.
\newblock Deep lagrangian networks: Using physics as model prior for deep
  learning.
\newblock {\em arXiv preprint arXiv:1907.04490}, 2019.

\bibitem{weinan2017proposal}
E~Weinan.
\newblock A proposal on machine learning via dynamical systems.
\newblock {\em Communications in Mathematics and Statistics}, 5(1):1--11, 2017.

\bibitem{lu2018beyond}
Yiping Lu, Aoxiao Zhong, Quanzheng Li, and Bin Dong.
\newblock Beyond finite layer neural networks: Bridging deep architectures and
  numerical differential equations.
\newblock In {\em International Conference on Machine Learning}, pages
  3276--3285. PMLR, 2018.

\bibitem{NEURIPS2019_42a6845a}
Yulia Rubanova, Ricky T.~Q. Chen, and David~K Duvenaud.
\newblock Latent ordinary differential equations for irregularly-sampled time
  series.
\newblock In {\em Advances in Neural Information Processing Systems},
  volume~32. Curran Associates, Inc., 2019.

\bibitem{dupont2019anode}
Emilien Dupont, Arnaud Doucet, and Yee~Whye Teh.
\newblock Augmented neural odes.
\newblock In {\em Advances in Neural Information Processing Systems},
  volume~32. Curran Associates, Inc., 2019.

\bibitem{gholami2019anode}
Amir Gholami, Kurt Keutzer, and George Biros.
\newblock Anode: Unconditionally accurate memory-efficient gradients for neural
  odes.
\newblock {\em arXiv preprint arXiv:1902.10298}, 2019.

\bibitem{zhuang2020adaptive}
Juntang Zhuang, Nicha Dvornek, Xiaoxiao Li, Sekhar Tatikonda, Xenophon
  Papademetris, and James Duncan.
\newblock Adaptive checkpoint adjoint method for gradient estimation in neural
  ode.
\newblock In {\em International Conference on Machine Learning}, pages
  11639--11649. PMLR, 2020.

\bibitem{zhuang2021mali}
Juntang Zhuang, Nicha~C Dvornek, Sekhar Tatikonda, and James~S Duncan.
\newblock Mali: A memory efficient and reverse accurate integrator for neural
  odes.
\newblock {\em arXiv preprint arXiv:2102.04668}, 2021.

\bibitem{zhang2019anodev2}
Tianjun Zhang, Zhewei Yao, Amir Gholami, Kurt Keutzer, Joseph Gonzalez, George
  Biros, and Michael Mahoney.
\newblock Anodev2: A coupled neural ode evolution framework.
\newblock {\em arXiv preprint arXiv:1906.04596}, 2019.

\bibitem{massaroli2020dissecting}
Stefano Massaroli, Michael Poli, Jinkyoo Park, Atsushi Yamashita, and Hajime
  Asma.
\newblock Dissecting neural odes.
\newblock In {\em 34th Conference on Neural Information Processing Systems,
  NeurIPS 2020}. The Neural Information Processing Systems, 2020.

\bibitem{quaglino2019snode}
Alessio Quaglino, Marco Gallieri, Jonathan Masci, and Jan Koutn{\'\i}k.
\newblock {SNODE}: Spectral discretization of neural odes for system
  identification.
\newblock {\em arXiv preprint arXiv:1906.07038}, 2019.

\bibitem{maulik2020time}
Romit Maulik, Arvind Mohan, Bethany Lusch, Sandeep Madireddy, Prasanna
  Balaprakash, and Daniel Livescu.
\newblock Time-series learning of latent-space dynamics for reduced-order model
  closure.
\newblock {\em Physica D: Nonlinear Phenomena}, 405:132368, 2020.

\bibitem{portwood2019turbulence}
Gavin~D Portwood, Peetak~P Mitra, Mateus~Dias Ribeiro, Tan~Minh Nguyen,
  Balasubramanya~T Nadiga, Juan~A Saenz, Michael Chertkov, Animesh Garg, Anima
  Anandkumar, Andreas Dengel, et~al.
\newblock Turbulence forecasting via neural ode.
\newblock {\em arXiv preprint arXiv:1911.05180}, 2019.

\bibitem{lee2020parameterized}
Kookjin Lee and Eric~J Parish.
\newblock Parameterized neural ordinary differential equations: Applications to
  computational physics problems.
\newblock {\em arXiv preprint arXiv:2010.14685}, 2020.

\bibitem{toth2019hamiltonian}
Peter Toth, Danilo~J Rezende, Andrew Jaegle, S{\'e}bastien Racani{\`e}re,
  Aleksandar Botev, and Irina Higgins.
\newblock Hamiltonian generative networks.
\newblock In {\em International Conference on Learning Representations}, 2019.

\bibitem{lutter2018deep}
Michael Lutter, Christian Ritter, and Jan Peters.
\newblock Deep lagrangian networks: Using physics as model prior for deep
  learning.
\newblock In {\em International Conference on Learning Representations}, 2018.

\bibitem{huh2020time}
In~Huh, Eunho Yang, Sung~Ju Hwang, and Jinwoo Shin.
\newblock Time-reversal symmetric ode network.
\newblock {\em Advances in Neural Information Processing Systems}, 33, 2020.

\bibitem{trask2020enforcing}
Nathaniel Trask, Andy Huang, and Xiaozhe Hu.
\newblock Enforcing exact physics in scientific machine learning: a data-driven
  exterior calculus on graphs.
\newblock {\em arXiv preprint arXiv:2012.11799}, 2020.

\bibitem{lee2019deep}
Kookjin Lee and Kevin Carlberg.
\newblock Deep conservation: A latent-dynamics model for exact satisfaction of
  physical conservation laws.
\newblock {\em arXiv preprint arXiv:1909.09754}, 2019.

\bibitem{beucler2019achieving}
Tom Beucler, Stephan Rasp, Michael Pritchard, and Pierre Gentine.
\newblock Achieving conservation of energy in neural network emulators for
  climate modeling.
\newblock {\em arXiv preprint arXiv:1906.06622}, 2019.

\bibitem{hernandez2021structure}
Quercus Hern{\'a}ndez, Alberto Bad{\'\i}as, David Gonz{\'a}lez, Francisco
  Chinesta, and El{\'\i}as Cueto.
\newblock Structure-preserving neural networks.
\newblock {\em Journal of Computational Physics}, 426:109950, 2021.

\bibitem{oettinger2014irreversible}
Hans~Christian Oettinger.
\newblock Irreversible dynamics, onsager-casimir symmetry, and an application
  to turbulence.
\newblock {\em Physical Review E}, 90(4):042121, 2014.

\bibitem{paszke2017automatic}
Adam Paszke, Sam Gross, Soumith Chintala, Gregory Chanan, Edward Yang, Zachary
  DeVito, Zeming Lin, Alban Desmaison, Luca Antiga, and Adam Lerer.
\newblock Automatic differentiation in pytorch.
\newblock 2017.

\bibitem{dormand1980family}
John~R Dormand and Peter~J Prince.
\newblock A family of embedded runge-kutta formulae.
\newblock {\em Journal of computational and applied mathematics}, 6(1):19--26,
  1980.

\bibitem{kingma2014adam}
Diederik~P Kingma and Jimmy Ba.
\newblock Adam: A method for stochastic optimization.
\newblock {\em arXiv preprint arXiv:1412.6980}, 2014.

\bibitem{shang2020structure}
Xiaocheng Shang and Hans~Christian {\"O}ttinger.
\newblock Structure-preserving integrators for dissipative systems based on
  reversible--irreversible splitting.
\newblock {\em Proceedings of the Royal Society A}, 476(2234):20190446, 2020.

\bibitem{kloeden1992stochastic}
Peter~E Kloeden and Eckhard Platen.
\newblock Stochastic differential equations.
\newblock In {\em Numerical Solution of Stochastic Differential Equations},
  pages 103--160. Springer, 1992.

\end{thebibliography}
}
\end{document}